\documentclass[12pt]{article} 

\usepackage{amsmath,amssymb}
\usepackage{graphics,epsfig,psfrag}
\usepackage{bigints}

\title{The shadow of a collapsing dark star}

\author{Stefanie Schneider and Volker Perlick
\\[0.3cm]
{\small ZARM, University of Bremen, 28359 Bremen, Germany} 
\\
{\small Email: perlick@zarm.uni-bremen.de}
}

\date{}

\begin{document}

\maketitle

\begin{abstract}
The shadow of a black hole is usually calculated, either analytically
or numerically, on the assumption that the black hole is eternal, i.e.,
that it has existed for all time.
Here we ask the question of how this shadow comes about in
the course of time when a black hole is formed by gravitational
collapse. To that end we consider a star that is spherically symmetric,
dark and non-transparent and we assume that it begins, at some
instant of time, to collapse in free fall like a ball of dust. We 
analytically calculate the dependence on time of the angular radius of
the shadow, first for a static observer who is watching the collapse
from a certain distance and then for an observer who is falling
towards the centre following the collapsing star. 
%\keywords{black hole \and shadow \and gravitational collapse \and Painlev{\'e}-Gullstrand coordinates}
% \PACS{PACS code1 \and PACS code2 \and more}
% \subclass{MSC code1 \and MSC code2 \and more}
\end{abstract}

%%%%%%%%%%%%%%%%%%%%%%%%%%%%%%%%%%%%
\section{Introduction}\label{sec:intro}

When a black hole is viewed against a backdrop of light sources,
the observer sees a black disc in the sky which is known as the
\emph{shadow} of the black hole. Points inside this black disc
correspond to past-oriented light rays that go from the observer
towards the horizon of the black hole, while points outside this
black disc correspond to past-oriented light rays that are more
or less deflected by the black hole and then meet one of the 
light sources. For a Schwarzschild black hole, which is non-rotating
the shadow is circular and its boundary corresponds to light rays 
that asymptotically spiral towards circular photon orbits
that fill the so-called photon sphere at 1.5 Schwarzschild radii
around the black hole. For a Kerr black hole, which is rotating, 
the shadow is flattened on one side and its boundary corresponds 
to light rays that spiral towards spherical photon orbits
that fill a 3-dimensional photon region around the black hole.
For the supermassive black hole that is assumed to sit at the
centre of our Galaxy, the predicted angular diameter of the shadow is 
about 53 microarcseconds which is within reach of VLBI observations.
There is an ongoing effort to actually observe this shadow, and 
also the one of the second-best black-hole candidate at the centre
of M87, see   {\tt http://www.eventhorizontelescope.org}.

When calculating the shadow one usually considers an \emph{eternal}
black hole, i.e., a black hole that is static or stationary and exists
for all time. For a Schwarzschild black hole, there is a simple
analytical formula for the angular radius of the shadow which goes
back to Synge \cite{Synge1966}. (Synge did not use the word ``shadow''
which was introduced much later. He calculated what he called the
\emph{escape cones} of light. The opening angle of the escape 
cone is the complement of the angular radius of the shadow.) For
a Kerr black hole, the shape of the shadow was 
calculated for an observer at infinity by Bardeen \cite{Bardeen1973}.
More generally, an analytical formula for the boundary curve
of the shadow was given, for an observer anywhere
in the domain of outer communication of a Pleba{\'n}ski-Demia{\'n}ski
black hole, by Grenzebach et al. 
\cite{GrenzebachPerlickLaemmerzahl2014,GrenzebachPerlickLaemmerzahl2015}.
For the Kerr case, this formula was further evaluated by Tsupko
\cite{Tsupko2017}.
These analytical results are complemented by ambitious numerical
studies, performing ray tracing in black hole spacetimes with various
optical effects taken into account. We mention in particular a 
paper by Falcke et al. \cite{FalckeMeliaAgol2000} where the perspectives
of actually observing black-hole shadows were numerically investigated
taking the presence of emission regions and scattering into account, and
a more recent article by James et al.   
\cite{JamesTunzelmannFranklinThorne2015}
which focusses on the numerical work that was done for the
movie \emph{Interstellar} but also reviews earlier work.

As we have already emphasised, in all these analytical and numerical
works an eternal black hole is considered. Actually, we believe that
black holes are not eternal: They have come into existence some finite
time ago by gravitational collapse (and are then possibly growing by 
accretion or mergers with other black holes). This brings us to the 
question of how an observer who is watching the collapse would see 
the shadow coming about in the course of time. This is the question 
we want to investigate in this paper.

The visual appearance of a star undergoing gravitational collapse
has been studied in several papers, beginning with the pioneering
work of Ames and Thorne \cite{AmesThorne1968}. In this work, and 
in follow-up papers e.g. by Jaffe \cite{Jaffe1969}, Lake and Roeder
\cite{LakeRoeder1979} and Frolov et al. \cite{FrolovKimLee2007},
the emphasis is on the frequency shift of light coming from the 
surface of the collapsing star. More recent papers by Kong et al.
\cite{KongMalafarinaBambi2014,KongMalafarinaBambi2015} and
by Ortiz et al. 
\cite{OrtizSarbachZannias2015a,OrtizSarbachZannias2015b}
investigated the frequency shift of light passing through a 
collapsing transparent star, thereby contrasting the collapse
to a black hole with the collapse to a naked singularity. In 
contrast to all these earlier articles, here we consider
a \emph{dark} and \emph{non-transparent} collapsing star which
is seen as a black disc when viewed against a backdrop of light 
sources and we ask how this black disc changes 
in the course of time. 

For the collapsing star we use a particularly simple model: We
assume that the star is spherically symmetric and that it begins
to collapse, at some instant of time, in free fall like a ball of
dust until it ends up in a point singularity at the centre. 
The metric inside such a collapsing ball of dust was
found in a classical paper by Oppenheimer and Snyder
\cite{OppenheimerSnyder1939}. However, for our purpose, as 
we assume the collapsing star to be non-transparent, we do
not need this interior metric. All we need to know is
that a point on the surface follows 
a timelike geodesic in the ambient Schwarzschild spacetime. 
We will demonstrate that in this situation the time 
dependence of the shadow can be given analytically. We do
this first for a static observer who is watching the collapse from a
certain distance, and then also for an observer who is falling
towards the centre and ending up in the point singularity 
after it has formed. The latter situation is (hopefully) not
of relevance for practical astronomical observations but
we believe that the calculation is quite instructive from
a conceptual point of view.

The paper is organised as follows.   In Section \ref{sec:PG}
we review some basic facts on the Schwarzschild solution 
in Painlev{\'e}-Gullstrand coordinates. These coordinates 
are particularly well suited for our purpose because they
are regular at the horizon, so they allow to consider worldlines
of observers or light signals that cross the horizon without the
need of patching different coordinate charts together. In 
Section \ref{sec:shbh} we rederive in Painlev{\'e}-Gullstrand
coordinates the equations for the shadow of an eternal 
Schwarzschild black hole. We do this both for a static and
for an infalling observer. The results of this section will then
be used in the following two sections for calculating the shadow 
of a collapsing star. We do this first for a static observer
in Section \ref{sec:shcoll1} and then for an infalling observer
in Section \ref{sec:shcoll2}. We summarise our results in 
Section \ref{sec:conclusions}. 

%%%%%%%%%%%%%%%%%%%%%%%%%%%%%%%%%%%%%%%%%%%%%%%%%%%%%%%%%%%%%%%%%%%%%%
\section{Schwarzschild metric in Painlev{\'e}-Gullstrand coordinates}\label{sec:PG}

Throughout this paper, we work with the Schwarzschild metric 
in Painlev{\'e}-Gullstrand coordinates \cite{Painleve1921,Gullstrand1922},

\begin{equation}\label{eq:gPG}
g_{\mu \nu}dx^{\mu} dx^{\nu} 
= - \left( 1 - \dfrac{2m}{r} \right) c^2 d T ^2 + 2 \sqrt{\dfrac{2m}{r}} \, c \, d T \, dr
+ dr^2 +
r^2 \big( d \vartheta ^2 + \mathrm{sin} ^2 \vartheta \, 
d \varphi ^2 \big) \, .
\end{equation}
Here

\begin{equation}\label{eq:m}
m=\dfrac{GM}{c^2}
\end{equation}
is the mass parameter with the dimension of a length; $M$ is 
the mass of the central object in SI units, $G$ is Newton's 
gravitational constant and $c$ is the vacuum speed of light.

The Painlev{\'e}-Gullstrand coordinates $(T,r,\vartheta, \varphi )$
are related to the standard text-book Schwarzschild coordinates
$(t,r,\vartheta , \varphi )$ by

\begin{equation}\label{eq:Tt}
c \, dT= c \, dt + 
\sqrt{\dfrac{2m}{r}} \, \dfrac{dr}{\Big( 1- \dfrac{2m}{r} \Big)}
\, .
\end{equation} 
As a historical side remark, we mention that both Painlev{\'e}
\cite{Painleve1921} and Gullstrand \cite{Gullstrand1922} believed
that they had found a new solution to Einstein's vacuum field
equation before Lema{\^\i}tre \cite{Lemaitre1933} demonstrated 
that it is just the Schwarzschild solution in other coordinates.  
Whereas in the standard Schwarzschild coordinates the metric
has a coordinate singularity at the horizon at $r=2m$, in the
Painlev{\'e}-Gullstrand coordinates the metric is regular
on the entire domain $0 < r < \infty$.

On the domain $2m<r<\infty$ we will use the tetrad
\begin{gather}
\nonumber
e_0 = \dfrac{1}{c \sqrt{1- \dfrac{2m}{r}}}
\, \dfrac{\partial}{\partial T} \, , \quad
e_1 = \sqrt{1-\dfrac{2m}{r}} \dfrac{\partial}{\partial r} +
\dfrac{\sqrt{\dfrac{2m}{r}}}{c \sqrt{1- \dfrac{2m}{r}}} \, \dfrac{\partial}{\partial T} \, , \quad
\\
e_2 = \dfrac{1}{r }
\, \dfrac{\partial}{\partial \vartheta} \, , \quad
e_3 = \dfrac{1}{r \, \mathrm{sin} \, \vartheta}
\, \dfrac{\partial}{\partial \varphi} \, .
\label{eq:emu}
\end{gather}
From (\ref{eq:gPG}) we read that this tetrad is orthonormal,
%\begin{equation}\label{eq:ortho}
$g _{\mu \nu} e_{\rho} ^{\mu} e_{\sigma} ^{\nu} 
= \eta _{\rho \sigma}$
%\end{equation}
with $(\eta _{\rho \sigma}) = \mathrm{diag} (-1,1,1,1)$,
for $2m<r<\infty$. Up to a factor of $c$, the vector field
$e_0$ is the four-velocity field of observers that stay at
fixed spatial coordinates $(r, \vartheta , \varphi )$. We
refer to them as to the \emph{static} observers.

From (\ref{eq:gPG}) we find that, for a static observer 
at radius coordinate $r(>2m)$, proper time $\tau$ is related 
to the Painlev{\'e}-Gullstrand time coordinate $T$ by 
\begin{equation}\label{eq:taustatic}
\dfrac{dT}{d \tau} = \dfrac{1}{\sqrt{1- \dfrac{2m}{r}}}
\, .
\end{equation}

In the Painlev{\'e}-Gullstrand coordinates, the geodesics 
in the Schwarzschild spacetime are the solutions to the
Euler-Lagrange equations of the Lagrangian 
 
\begin{equation}\label{eq:LagPG}
\mathcal{L} \big( x , \dot{x} \big) 
= \dfrac{1}{2} \left(
- \left( 1 - \dfrac{2m}{r} \right) c^2 \dot{T}{} ^2 + 
2 \sqrt{\dfrac{2m}{r}} \, c \, \dot{T} \, \dot{r}
+ \dot{r}{}^2 +
r^2  \, \Big( \mathrm{sin}^2 \vartheta \, \dot{\varphi}{} ^2 
+ \dot{\vartheta}{} \Big) \right) 
\, .
\end{equation}
Here the overdot means derivative with respect to an affine 
parameter.

The $T$ and $\varphi$ components of the Euler-Lagrange
equations give us the familiar constants of motion 
$E$ and $L$ in Painlev{\'e}-Gullstrand coordinates,

\begin{equation}\label{eq:E}
E = - \dfrac{\partial \mathcal{L}}{\partial \dot{T}} 
= \left( 1 - \dfrac{2m}{r} \right) c^2 \dot{T} 
- \sqrt{\dfrac{2m}{r}} \, c \, \dot{r} 
\end{equation}
and

\begin{equation}\label{eq:L}
L =  
\dfrac{\partial \mathcal{L}}{\partial \dot{\varphi}} 
= r^2 \mathrm{sin} ^2 \vartheta \, \dot{\varphi} \, .
\end{equation}
For the purpose of this paper we will need the 
radial timelike geodesics and the lightlike geodesics
in the equatorial plane. 

%-------------------------------------------------------------------------------------------------------------------------------
\subsection{Radial timelike geodesics}\label{sec:radial}

We consider massive objects in radial free 
fall, i.e., radial geodesics ($\dot{\varphi}=0$ and 
$\dot{\vartheta} =0$) which are timelike. Then we may 
choose the affine parameter equal to proper time $\tau$, 

\begin{equation}\label{eq:Lagrad}
- \, c^2 =
 - \left( 1 - \dfrac{2m}{r} \right) c^2 \Big( \dfrac{dT}{d \tau} \Big) ^2 
+ 2 \sqrt{\dfrac{2m}{r}} \, c \, \dfrac{dT}{d \tau} \, \dfrac{dr}{d \tau}
+ \Big( \dfrac{dr}{d \tau} \Big) ^2 \, .
\end{equation}
In this notation (\ref{eq:E}) can be rewritten as

\begin{equation}\label{eq:Erad}
\varepsilon := \dfrac{E}{c^2} =  
\left( 1 - \dfrac{2m}{r} \right) \dfrac{dT}{d \tau} 
- \sqrt{\dfrac{2m}{r}} \, \dfrac{1}{c} \, \dfrac{dr}{d \tau} 
\end{equation}
whereas (\ref{eq:L}) requires $L=0$. 

\begin{figure}[h]
    \psfrag{x}{$r$} % 
    \psfrag{y}{$cT$} % 
    \psfrag{a}{\small $\,$ \hspace{-0.6cm} $\begin{matrix} \, \\[-0.17cm] 2 m\end{matrix}$} % 
    \psfrag{b}{\small $\,$ \hspace{-0.6cm} $\begin{matrix} \, \\[-0.2cm] 10 m\end{matrix}$} % 
    \psfrag{c}{\small $\,$ \hspace{-0.6cm} $\begin{matrix} \, \\[-0.2cm] 20 m\end{matrix}$} % 
    \psfrag{d}{\small $\,$ \hspace{-0.6cm} $\begin{matrix} \, \\[-0.2cm] 30 m\end{matrix}$} % 
\centerline{\epsfig{figure=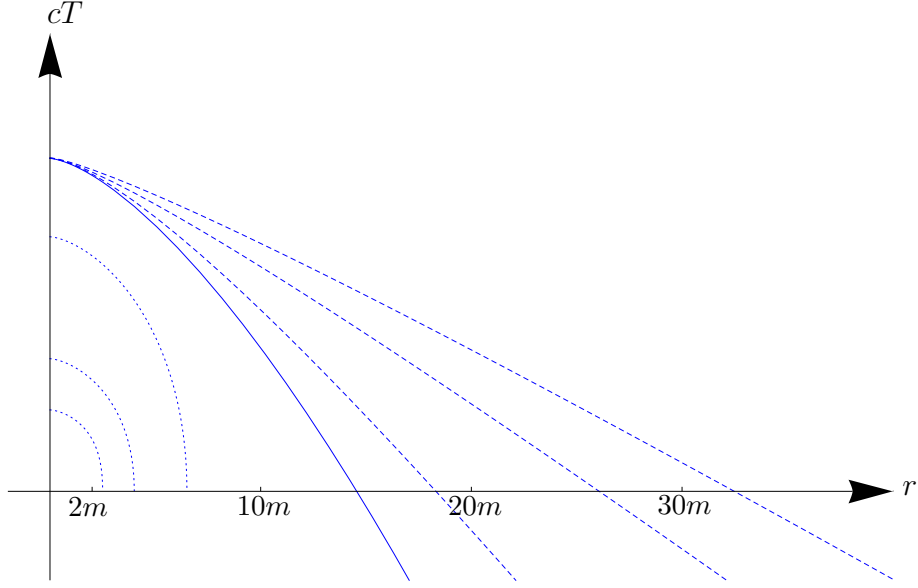, width=12cm}}
\caption{Radially infalling observers with $\varepsilon <1$ (dotted), 
$\varepsilon =1$ (solid) and $\varepsilon >1$ (dashed). For the four 
worldlines with $\varepsilon \ge 1$ we have chosen the initial conditions 
such that they arrive simultaneously at $r=0$. For the three worldlines 
with $\varepsilon <1$ we have assumed that they start from rest at the 
same Painlev{\'e}-Gullstrand time; then these worldlines do not arrive 
simultaneously at $r=0$. As an alternative, one could have considered 
worldlines that start from rest at the same Schwarzschild time; they 
would arrive simultaneously at $r=0$.
\label{fig:radial}}
\end{figure}

In the following we restrict to the case that the parametrisation by proper
time is future oriented with respect to the Painlev{\'e}-Gullstrand
time coordinate, $dT/d \tau >0$, and we consider only ingoing motion,
$dr/d \tau <0$, that starts in the domain $r>2m$. Then $\varepsilon >0$ 
and (\ref{eq:Erad}) and (\ref{eq:Lagrad}) imply

\begin{equation}\label{eq:rTrad}
\dfrac{dr}{d \tau} = 
- \, c \, \sqrt{\varepsilon ^2- 1+ \dfrac{2m}{r}} 
\, , \quad
\dfrac{dT}{d \tau} =
\dfrac{
\varepsilon  - \sqrt{\dfrac{2m}{r}}   
\sqrt{\varepsilon ^2- 1 + \dfrac{2m}{r} }
 }{
1 - \dfrac{2m}{r}  
}
\end{equation} 
We distinguish three cases, see Fig.~\ref{fig:radial}:

(a) $0<\varepsilon < 1$: Then $dr/d \tau =0$ at a radius coordinate 
$r_i$ given by $\varepsilon ^2=1-2m/r_i$, i.e., this case 
describes free fall from rest at $r_i$. Clearly, the 
possible values of $r_i$ are $2m<r_i<\infty$.

(b) $\varepsilon =1$: This is the limit of case (a) for 
$r_i \to \infty$, i.e., free fall from rest at infinity. 
It is usual to refer to such freely falling observers as to
the \emph{Painlev{\'e}-Gullstrand observers}. In this case 
the two equations (\ref{eq:rTrad}) reduce to 

\begin{equation}\label{eq:PGeom}
\dfrac{dr}{d \tau} = - \, c \, \sqrt{\dfrac{2m}{r}} 
\, , \quad 
\dfrac{dT}{d \tau} = 1 \, .
\end{equation}
The second equation shows that the coordinate $T$ gives proper 
time along the worldlines of the Painlev{\'e}-Gullstrand 
observers .

(c) $1 < \varepsilon < \infty$: These are freely falling observers that come in 
from infinity with a non-zero inwards-directed initial velocity 
and then fall towards the centre.

Choosing a value of $\varepsilon >0$ defines a family of infalling observers.
We associate with this family the tetrad 

\begin{gather}
\nonumber
\tilde{e}{}_0 =
\dfrac{
\varepsilon -\sqrt{\dfrac{2m}{r}} \sqrt{\varepsilon ^2-1+\dfrac{2m}{r}}
}{ 
\Big( 1 - \dfrac{2m}{r} \Big) \, c
}
\, \dfrac{\partial}{\partial T} 
-
\sqrt{\varepsilon ^2-1+\dfrac{2m}{r}} \, \dfrac{\partial }{\partial r} \, ,
\\
\nonumber
\tilde{e}{}_1 = \varepsilon \, \dfrac{\partial}{\partial r} 
+
\dfrac{
\varepsilon \sqrt{\dfrac{2m}{r}} -\sqrt{\varepsilon ^2-1+\dfrac{2m}{r}}
}{
\Big( 1- \dfrac{2m}{r} \Big) \, c
}
\dfrac{\partial}{\partial T} \, ,
\\
\tilde{e}{}_2 = \dfrac{1}{r }
\, \dfrac{\partial}{\partial \vartheta} \, , \quad
\tilde{e}{}_3 = \dfrac{1}{r \, \mathrm{sin} \, \vartheta}
\, \dfrac{\partial}{\partial \varphi} \, .
\label{eq:temu}
\end{gather}
For $\varepsilon \ge 1$ this tetrad is well-defined and orthonormal
on the entire domain $0 < r < \infty$. For $0 < \varepsilon < 1$ the tetrad
is restricted to the domain $0 < r < r_i = 2m/(1- \varepsilon ^2)$.

The relative velocity $v$ of a radially infalling observer 
with respect to the static observer at the same event can 
be calculated from the special relativistic formula

\begin{equation}\label{eq:v1}
g_{\mu \nu} \, e_0^{\mu} \, \tilde{e}{}_0^{\nu} 
= \dfrac{-1}{\sqrt{1-\dfrac{v^2}{c^2}}} \, .
\end{equation}
This results in

\begin{equation}\label{eq:vrad}
\dfrac{v}{c} = 
\dfrac{1}{\varepsilon} \, \sqrt{\varepsilon ^2- 1+ \dfrac{2m}{r} }
\, .
\end{equation}
Clearly, this formula makes sense only on the domain on which 
both families of observers are defined. For $\varepsilon \ge 1$ this
is true on the domain $2m<r<\infty$ whereas for $0< \varepsilon < 1$
it is true on the domain $2m < r < r_i = 2m/(1- \varepsilon ^2)$.

%-------------------------------------------------------------------------------------------------------------------------------
\subsection{Lightlike geodesics in the equatorial plane}\label{sec:light}

We will now rederive some results on lightlike geodesics
in the Schwarzschild spacetime, using Painlev{\'e}-Gullstrand 
coordinates. Because of spherical symmetry, it suffices
to consider geodesics in the equatorial plane, $\vartheta = \pi /2$. For
lightlike geodesics the Lagrangian is equal to zero,

\begin{equation}\label{eq:light}
0 = - \left( 1 - \dfrac{2m}{r} \right) c^2 \dot{T}{} ^2 
+ 2 \sqrt{\dfrac{2m}{r}} \, c \, \dot{T} \, \dot{r}
+ \dot{r}{}^2 +
r^2 \, \dot{\varphi}{}^2  \, .
\end{equation}
Dividing by $\dot{\varphi}{}^2$ and using (\ref{eq:E}) and 
(\ref{eq:L}) yields

\begin{equation}\label{eq:orbit}
\dfrac{dr}{d \varphi}  = 
\dfrac{\dot{r}}{\dot{\varphi}}
=
\pm \sqrt{\dfrac{E^2r^4}{c^2L^2} - r^2 +2mr}
\end{equation}
Reinserting this result into (\ref{eq:light}) gives us the 
equation for the Painlev{\'e}-Gullstrand travel time of light,

\begin{equation}\label{eq:trtime}
c \, \dfrac{dT}{d r}  = 
c \, \dfrac{\dot{T}}{\dot{r}}
= 
\dfrac{\sqrt{2mr}}{r-2m}  \pm 
\dfrac{
r
}{
(r-2m)\sqrt{1- ( r - 2m ) \dfrac{c^2L^2}{E^2r^3}}
} \, .
\end{equation}
For all $r>2m$, in the last expression the second term is 
bigger than the first. Therefore, in this domain the 
upper sign has to be chosen if $dT/dr > 0$  and the 
lower sign has to be chosen if $dT/dr <0$.

By differentiating (\ref{eq:orbit}) with respect to $\varphi$ 
we find 

\begin{equation}\label{eq:d2rdphi}
\dfrac{d^2r}{d \varphi ^2} =
\dfrac{4E^2r^3}{c^2L^2} - 2r +2m \, .
\end{equation}
If along a lightlike geodesic the radius coordinate goes 
through an extremum at value $r_m$, (\ref{eq:orbit}) and 
(\ref{eq:d2rdphi}) imply

\begin{equation}\label{eq:Erm}
0 = 
\dfrac{E^2r_m^3}{c^2L^2} - r_m +2m \, ,
\end{equation}

\begin{equation}\label{eq:Erm2}
\dfrac{d^2 r}{d \varphi ^2} \Big| _{r=r_m} = 
r_m-3m \, .
\end{equation}
This demonstrates that only local minima may occur in the domain $r>3m$  
and only local maxima may occur in the domain $r<3m$. The sphere at
$r=3m$ is filled with circular photon orbits that are unstable with respect 
to radial perturbations. These well-known facts will be crucial for the 
following analysis.

For a lightlike geodesic with an extremum of the radius coordinate at $r_m$,
(\ref{eq:Erm}) may be used for expressing $E^2/L^2$ in (\ref{eq:orbit}) and 
(\ref{eq:trtime}) in terms of $r_m$. This results in

\begin{equation}\label{eq:orbit2}
\dfrac{dr}{d \varphi}  = \pm \sqrt{\dfrac{(r_m-2m)r^4}{r_m^3} - r^2 +2mr}
\end{equation}
and

\begin{equation}\label{eq:trtime2}
c \, \dfrac{dT}{d r}  = 
\dfrac{\sqrt{2mr}}{(r-2m)}  \pm 
\dfrac{\sqrt{(r_m-2m)r^5}}{(r-2m) \sqrt{(r_m-2m)r^3-(r-2m)r_m^3}}
\end{equation}

\vspace{0.5cm}

%%%%%%%%%%%%%%%%%%%%%%%%%%%%%%%%%%%%%%%%%%%%%%%%%%%%%%%%%%%%%%%%%%%%%%
\section{The shadow of an eternal Schwarzschild black hole}\label{sec:shbh}
In this section we rederive the formulas for the angular radius
of the shadow of an eternal Schwarzschild black hole, both for
a static and for an infalling observer, using Painlev{\'e}-Gullstrand
coordinates. The results of this section will then be used for calculating 
the shadow of a collapsing star in the following sections. 

We consider a lightlike geodesic $\big( T(s),r(s),\varphi (s) \big)$ 
in the equatorial plane, where $s$ is an affine parameter. As before,
we denote the derivative with respect to $s$ by an overdot. We may
then expand the tangent vector of the lightlike geodesic with respect 
to the static tetrad (\ref{eq:emu}) and also, as an
alternative, with respect to the infalling tetrad 
(\ref{eq:temu}) for some chosen $\varepsilon >0$. Of course, 
the resulting equations are restricted to the domain where
the respective tetrad is well-defined and orthonormal. As the 
tangent vector is lightlike, these expansions may be written in 
terms of two angles $\alpha$ and $\tilde{\alpha}$,
\begin{gather}
\nonumber
\dot{T} \dfrac{\partial}{\partial T}
+
\dot{r} \dfrac{\partial}{\partial r}
+
\dot{\varphi} \dfrac{\partial}{\partial \varphi}
=
\chi \Big( 
e_0 + \mathrm{cos} \, \alpha \, e_1 -
\mathrm{sin} \, \alpha \, e_3 \Big) 
\\
=
\tilde{\chi} \Big( 
\tilde{e}{}_0 + \mathrm{cos} \, \tilde{\alpha} \, \tilde{e}{}_1 -
\mathrm{sin} \, \tilde{\alpha} \, \tilde{e}{}_3 \Big) \, .
\label{eq:tangent1}
\end{gather}
If the parametrisation of the lightlike geodesic is future-oriented
with respect to the $T$ coordinate, the scalar factors $\chi$ and 
$\tilde{\chi}$ are positive; otherwise they are negative.
$\alpha$ is the angle between the lightlike geodesic and 
the radial direction in the rest system of the static observer, 
whereas $\tilde{\alpha}$ is the analogously defined angle in 
the rest system of the infalling observer. 
$\alpha$ and $\tilde{\alpha}$ may take all values between 0 and $\pi$.
Of course, $\alpha$ is well-defined on the domain where the 
static observer exists (i.e., for $2m < r < \infty$) whereas
$\tilde{\alpha}$ is well-defined on the domain where the 
infalling observer exists (i.e., for $0 < r < r_i = 2m/(1-\varepsilon ^2)$ 
if $0< \varepsilon < 1$, and for $0 < r < \infty$ if $1 \le \varepsilon < \infty$).

Comparing coefficients of  $\partial/ \partial r$
and $\partial/\partial \varphi$ in (\ref{eq:tangent1}) yields

\begin{equation}\label{eq:cc2}
\dot{r} 
= \,  \chi \, \sqrt{1-\dfrac{2m}{r}} \, \mathrm{cos} \, \alpha  
= \tilde{\chi} \left( \varepsilon \, \mathrm{cos} \, \tilde{\alpha} 
- \sqrt{\varepsilon ^2 -1+\dfrac{2m}{r}} \right) \, ,
\end{equation}

\begin{equation}\label{eq:cc3}
\dot{\varphi} 
= - \chi \, \dfrac{\mathrm{sin} \, \alpha}{r} 
= - \tilde{\chi}  \, \dfrac{\mathrm{sin} \, \tilde{\alpha}}{r} \, .
\end{equation}

From (\ref{eq:cc2})  and (\ref{eq:cc3}) we find

\begin{equation}\label{eq:alpha1}
\dfrac{\dot{\varphi}}{\dot{r}} 
=
\dfrac{
- \mathrm{sin} \, \alpha
}{
r \sqrt{1-\dfrac{2m}{r}} \mathrm{cos} \, \alpha
}
=
\dfrac{
- \mathrm{sin} \, \tilde{\alpha}
}{
r \left( \varepsilon \, \mathrm{cos} \, \tilde{\alpha} -  
\sqrt{\varepsilon ^2 -1+\dfrac{2m}{r}} \right)
} \, .
\end{equation}
Now we apply these results to the case of a lightlike geodesic that goes through an extremum
of the radius coordinate at some value $r_m$. If we evaluate (\ref{eq:alpha1})
at a radius value $r>3m$ this extremum is necessarily a local minimum, whereas it is necessarily 
a local maximum if we evaluate (\ref{eq:alpha1}) at a radius value $r<3m$. 
In either case (\ref{eq:orbit2}) implies that the angles $\alpha$ and $\tilde{\alpha}$ 
at $r$ satisfy

\begin{equation}\label{eq:alpha2}
\dfrac{1}{\dfrac{r^2(r_m-2m)}{r_m^3}-1 + \dfrac{2m}{r}} 
=
\dfrac{ \mathrm{sin} ^2 \alpha}{ \left( 1-\dfrac{2m}{r} \right) 
\mathrm{cos} ^2 \alpha}
=
\dfrac{
\mathrm{sin} ^2 \tilde{\alpha}
}{
\left( \varepsilon \, \mathrm{cos} \, \tilde{\alpha} 
-  \sqrt{\varepsilon ^2 -1+\dfrac{2m}{r}} \right) ^2
} \, .
\end{equation}
From the second equality sign in (\ref{eq:alpha2}) we find

\begin{equation}\label{eq:aberr2}
\mathrm{sin} \, \alpha =
\dfrac{
\sqrt{1-\dfrac{2m}{r}} \, \dfrac{1}{\varepsilon} \, \mathrm{sin} \, \tilde{\alpha}
}{
1- \dfrac{1}{\varepsilon}  \sqrt{\varepsilon ^2 -1+\dfrac{2m}{r}} \, 
\mathrm{cos} \, \tilde{\alpha}
} \, .
\end{equation}
By (\ref{eq:vrad}), this  just demonstrates that $\alpha$ and $\tilde{\alpha}$ are 
related by the standard aberration formula. 

From the first equality sign in (\ref{eq:alpha2}) we find

\begin{equation}\label{eq:alpha4}
\mathrm{sin} \, \alpha =
\sqrt{\dfrac{r_m^3(r-2m)}{r^3(r_m-2m)}}
\end{equation}
and equating the first to the third expression in (\ref{eq:alpha2}) yields

\begin{equation}\label{eq:alpha5}
\mathrm{sin} \, \tilde{\alpha} =
\dfrac{
\sqrt{1 -\dfrac{2m}{r}} \,  
\sqrt{\dfrac{(r-2m)r_m^3}{(r_m-2m) r^3}} 
}{
\varepsilon \pm \, 
\sqrt{\varepsilon ^2 -1+\dfrac{2m}{r}} \, 
\sqrt{1-\dfrac{(r-2m)r_m^3}{(r_m-2m) r^3}}
} \, .
\end{equation}
In (\ref{eq:alpha5}) the upper sign is valid if $dr/d \varphi >0$ 
and the lower sign is valid if $dr/d \varphi <0$ at $r$.

\begin{figure}[h]
    \psfrag{x}{$r_O$} % 
    \psfrag{y}{$\,$ \hspace{-0.3cm} $\alpha _{\mathrm{sh}}$} % 
    \psfrag{a}{\small $\,$ \hspace{-0.5cm} $2m$} % 
    \psfrag{b}{\small $\,$ \hspace{-0.42cm} $3m$} % 
    \psfrag{c}{\small $\,$ \hspace{-0.5cm} $\dfrac{\pi}{2}$} % 
    \psfrag{d}{\small $\,$ \hspace{-0.5cm} $\pi$} % 
\centerline{\epsfig{figure=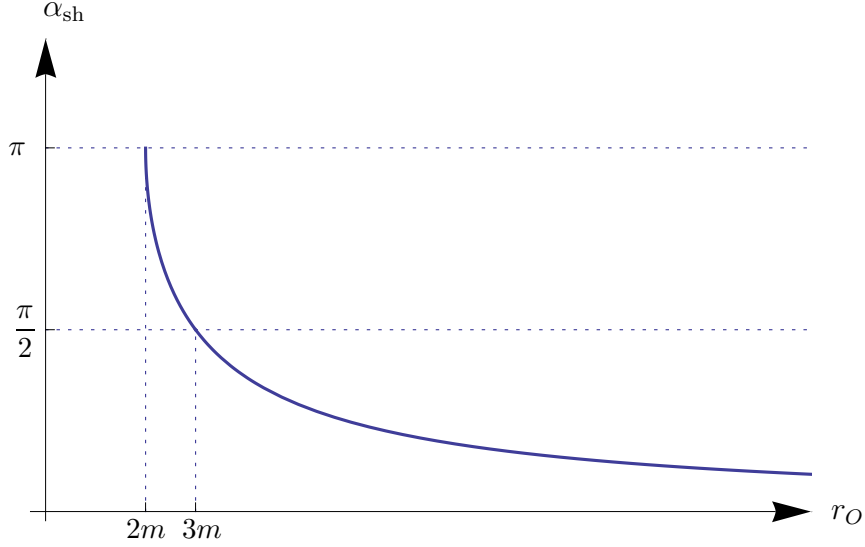, width=11cm}}
\caption{Angular radius $\alpha _{\mathrm{sh}}$ of the shadow of 
a Schwarzschild black hole for a static observer. What we have 
plotted here is Synge's formula (\protect{\ref{eq:alpha6}}). For 
an observer at $r_O=3m$, we have $\alpha = \pi/2$, i.e., half
of the sky is dark. For $r_O \to 2m$, we have $\alpha \to \pi$, 
i.e., in this limit the entire sky becomes dark. \label{fig:alphabh}}
\end{figure}

From (\ref{eq:alpha4}) and (\ref{eq:alpha5}) we can now easily determine 
the angular radius of the shadow. The latter is defined in the following way:
Consider an observer at radius coordinate $r=r_O >3m$. Then a lightlike geodesic 
issuing from the observer position into the past may either go to infinity,
possibly after passing through a minimum of the radius coordinate
at some $r_m>3m$, or it may go to the horizon. Similarly, for an observer position 
$2m<r_O<3m$ there are lightlike geodesics that go to the horizon, possibly
after passing through a maximum of the radius coordinate at some
$r_m<3m$, and lightlike geodesics that go to infinity. In either case 
the borderline between the two classes consists of lightlike geodesics 
that asymptotically spiral towards a circular lighlike geodesic at $r=3m$.
If we assume that there are light sources distributed in the spacetime 
anywhere but not between the observer and the black hole, then we have 
to associate darkness with the initial directions of lightlike geodesics
that go to the horizon and brightness with those that go to infinity.
This results in a circular black disc in the sky which is called the 
shadow of the black hole. The boundary of the shadow corresponds to 
lightlike geodesics that spiral towards $r=3m$.
\begin{figure}
    \psfrag{x}{$r_O$} % 
    \psfrag{y}{$\,$ \hspace{-0.3cm} $\tilde{\alpha}{}_{\mathrm{sh}}$} % 
    \psfrag{a}{\small $\,$ \hspace{-0.5cm} $\begin{matrix} \, \\[-0.17cm] 2 m\end{matrix}$} % 
    \psfrag{b}{\small $\,$ \hspace{-0.4cm} $\begin{matrix} \, \\[-0.17cm] 3 m\end{matrix}$} % 
    \psfrag{c}{\small $\,$ \hspace{-0.5cm} $\dfrac{\pi}{2}$} % 
    \psfrag{d}{\small $\,$ \hspace{-0.5cm} $\pi$} % 
\centerline{\epsfig{figure=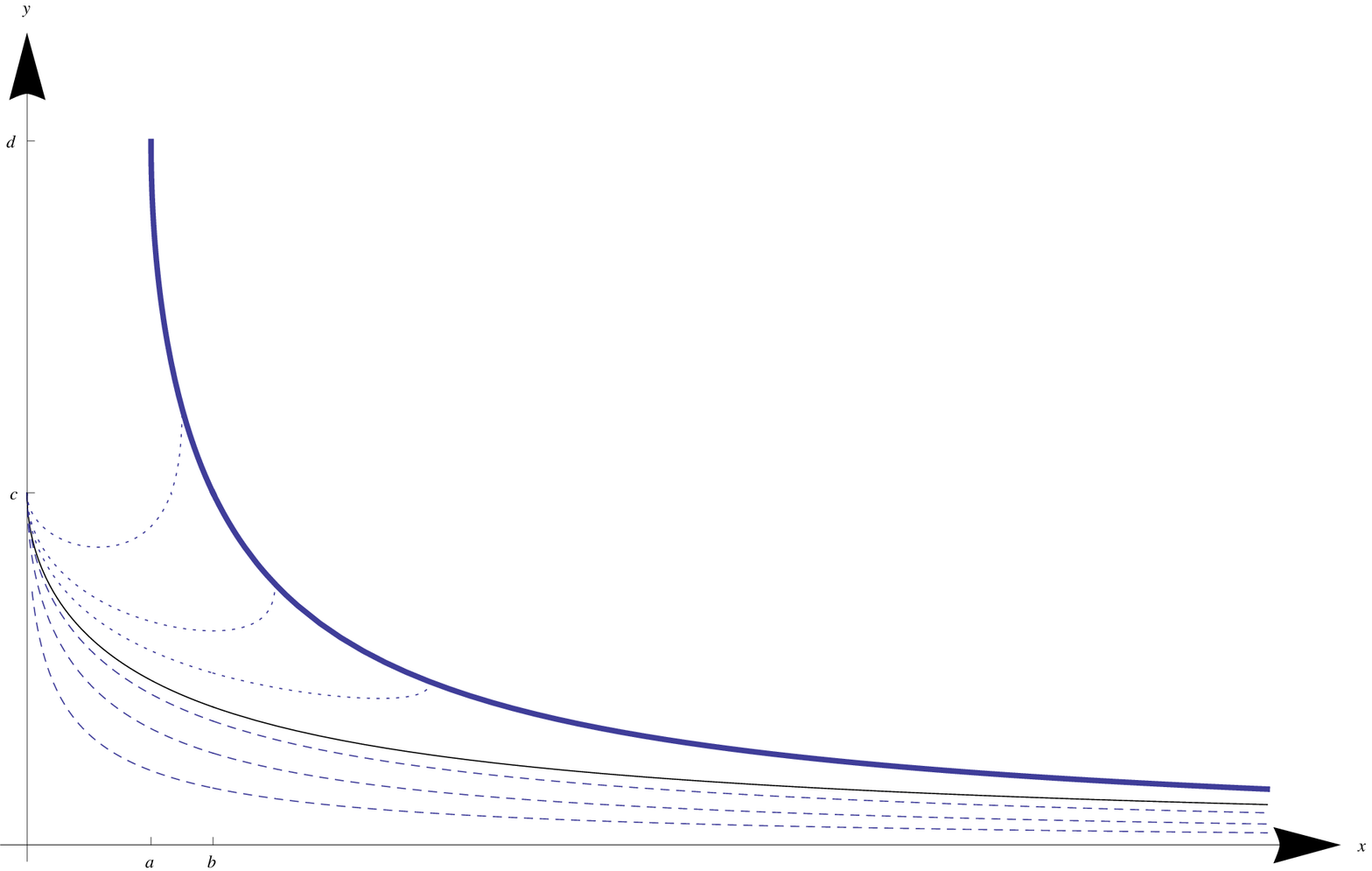, width=11.2cm}}
\caption{Angular radius $\tilde{\alpha}{}_{\mathrm{sh}}$ of the shadow 
of a Schwarzschild black hole for infalling observers with 
$\varepsilon < 1$ (dotted), $\varepsilon = 1$ (solid) and $\varepsilon > 1$ 
(dashed). In any case $\tilde{\alpha}{}_{\mathrm{sh}}$ approaches $\pi/2$, i.e., half of 
the sky becomes dark, for $r_O \to 0$. For the sake of comparison the 
angular radius $\alpha _{\mathrm{sh}}$ for a static observer is again
plotted in this diagram as a thick (blue) line, cf. 
Figure~\protect{\ref{fig:alphabh}}. \label{fig:talphabh}}
\end{figure}
Therefore, we get the angular radius of the shadow for an observer
at $r=r_O$ if we send $r_m \to 3m$ in (\ref{eq:alpha4}) and 
(\ref{eq:alpha5}). This results in 

\begin{equation}\label{eq:alpha6}
\mathrm{sin} \, \alpha _{\mathrm{sh}}=
\dfrac{\sqrt{27} m}{r_O} \sqrt{ 1- \dfrac{2m}{r_O}}
\end{equation}
and

\begin{equation}\label{eq:alpha7}
\mathrm{sin} \, \tilde{\alpha} {}_{\mathrm{sh}} =
\dfrac{
\dfrac{\sqrt{27} m}{r_O} \,  
\left(
\varepsilon  \mp  
\sqrt{\varepsilon ^2 -1+\dfrac{2m}{r_O}}
\, \sqrt{1- \dfrac{27 m^2}{r_O^2} 
\Big( 1 - \dfrac{2m}{r_O} \Big)} 
\right)
}{
1 + 27 \, \dfrac{m^2}{r_O^2} \, 
\left( \varepsilon ^2 -1+\dfrac{2m}{r_O} \right)
} \, ,
\end{equation}
respectively. (\ref{eq:alpha6}) gives us the angular radius 
$\alpha _{\mathrm{sh}}$ as it is seen by a static observer 
at $r_O$, see Figure~\ref{fig:alphabh}. 
This formula is known since Synge \cite{Synge1966}. It is 
meaningful only for observer positions $2m < r_O < \infty$ 
because the static observers exist on this domain only. 
By contrast, (\ref{eq:alpha7}) gives us the angular radius 
$\tilde{\alpha}{}_{\mathrm{sh}}$ of the shadow as it is 
seen by an infalling observer at momentary radius coordinate 
$r_O$, see Figure~\ref{fig:talphabh}. A similar formula was
derived by Bakala et al. \cite{BakalaEtAl2007}, even for 
the more general case of a Schwarzschild-deSitter (Kottler) 
black hole. (\ref{eq:alpha7}) is meaningful for 
$0<r_O<r_i=2m/(1-\varepsilon ^2)$ if $0 < \varepsilon < 1$
and for $0 < r_O < \infty$ if $1 \le \varepsilon < \infty$.
We have to choose the upper sign 
for $3m<r_O<\infty$ and the lower sign for $0 < r_O < 3m$. 
Nothing particular happens if the infalling observer crosses 
$r_O=3m$ or $r_O=2m$,
\begin{equation}\label{eq:2mrm}
\mathrm{sin} \, \tilde{\alpha} {}_{\mathrm{sh}} \Big| _{r_O=3m} 
= \dfrac{1}{\sqrt{3} \varepsilon} 
\, , \quad
\mathrm{sin} \, \tilde{\alpha} {}_{\mathrm{sh}} \Big| _{r_O=2m} 
= \dfrac{
\sqrt{27} \, \varepsilon
}{
1+\dfrac{27}{4} \, \varepsilon ^2
} \, .
\end{equation}
For calculating the limit $r_O \to 0$ we have to use 
(\ref{eq:alpha7}) with the lower sign. After multiplying 
numerator and denominator with $r_O^3$ we find that 
$\mathrm{sin} \, \tilde{\alpha} _{\mathrm{sh}} \to 1$, 
i.e., $\tilde{\alpha} _{\mathrm{sh}} \to \pi /2$. This 
shows that, independently of $\varepsilon$, the shadow covers 
half of the sky at the moment when the infalling observer 
ends up in the singularity in the centre, see again 
Figure~\ref{fig:talphabh} and cf. 
Bakala et al. \cite{BakalaEtAl2007}.
Expressing $r_O$ in terms of $\tau$ with the help of 
(\ref{eq:rTrad}) on the right-hand side of 
(\ref{eq:alpha5}) gives us $\tilde{\alpha}{}_{\mathrm{sh}}$ 
as a function of proper time $\tau$ of the infalling observer. 
 
\vspace{0.5cm}

%%%%%%%%%%%%%%%%%%%%%%%%%%%%%%%%%%%%%%%%%%%%%%%%%%%%%%%%%%%%%%%%%%%%%%
\section{The shadow of a collapsing star for a static observer}
\label{sec:shcoll1}

In this section we consider a spherically symmetric star that undergoes 
gravitational collapse and a static observer who is watching the collapse. 
The star is assumed to be dark and non-transparent. In analogy to
the black-hole case, we assume that there are light sources distributed
anywhere in the spacetime but not in the region between the observer 
and the star. By the latter we mean the region covered by past-oriented 
light rays from the observer that reach the surface of the star (before the
black hole has formed) or go to the horizon (after the black hole has
formed). Under these assumptions  the star will cast a circular shadow 
on the observer's sky. It is our goal to determine the angular radius of 
this shadow as a function of time.

\begin{figure}[h]
    \psfrag{x}{$r $} % 
    \psfrag{y}{$\,$ \hspace{-0.3cm} $T$} % 
    \psfrag{z}{\small $\,$ \hspace{-0.45cm} $\begin{matrix} \, \\[-0.2cm] r_O \end{matrix}$} % 
    \psfrag{a}{\small $\,$ \hspace{-0.35cm} $\begin{matrix} \, \\[-0.18cm] r_i \end{matrix}$} % 
    \psfrag{b}{\small $\,$ \hspace{-0.6cm} $\begin{matrix} \, \\[-0.2cm] r_S^{(2)} \end{matrix}$} % 
    \psfrag{p}{\small $\,$ \hspace{-0.8cm} $T_S^{(2)}$} % 
    \psfrag{q}{\small $\,$ \hspace{-0.9cm} $T_S^{\mathrm{coll}}$} % 
\centerline{\epsfig{figure=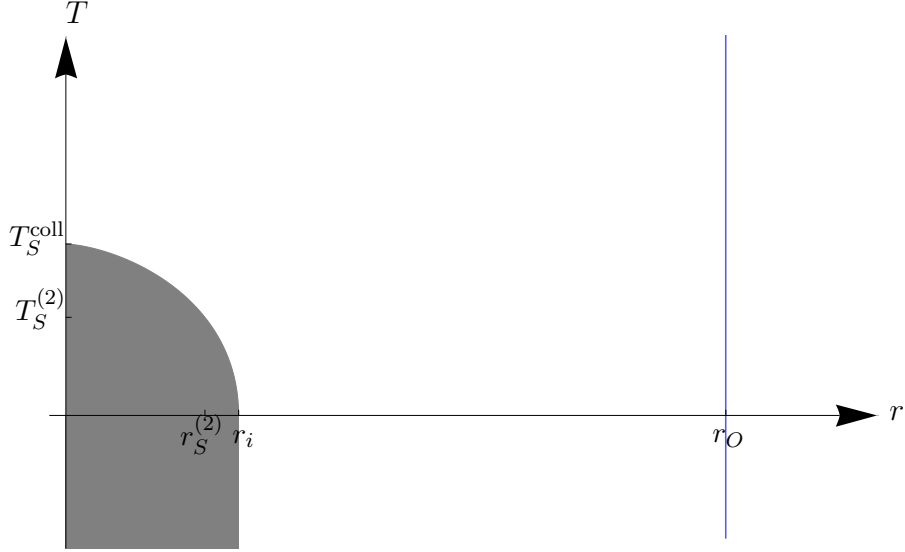, width=11.3cm}}
\caption{Spacetime diagram of a collapsing star and a static observer. 
The star begins to collapse at Painlev{\'e}-Gullstrand time
$T=0$ with radius $r_i$. The observer is static at radius $r_O$. \label{fig:collstat}}
\end{figure}

For the collapsing star we use the simplest model: We assume that the star
has constant radius $r_S=r_i$ up to Painlev{\'e}-Gullstrand time $T_S=0$ and then
collapses in free fall like a ball of dust, i.e., such that each point on the
surface of the star follows a radial timelike geodesic. Here and in the following we use the
index $S$ for the Painlev{\'e}-Gullstrand coordinates of the surface of the star, i.e., 
the star has radius $r_S$ at time $T_S$. For times $T_S>0$ the worldline of an
observer on the surface of the star is then  given by one of the dotted lines
in Figure~\ref{fig:radial}. From (\ref{eq:rTrad}) with $\varepsilon  ^2=1-2m/r_i$
we find that for $T_S>0$

\begin{equation}\label{eq:surface}
c T_S  = 
\bigintss _{r_S} ^{r_i} 
\dfrac{  \sqrt{r_i-2m} \, \sqrt{r^3} \, dr}{(r-2m) \sqrt{2m(r_i-r)}}
- \bigintss _{r_S}^{r_i} \dfrac{\sqrt{2mr} \, dr}{(r-2m)}
\end{equation}
Equating $r_S$ to zero gives the collapse time, $T_S^{\mathrm{coll}}$, i.e., the
Painlev{\'e}-Gullstrand time when the star has collapsed to a point singularity 
at the centre, see Figure~\ref{fig:collstat},

\begin{equation}\label{eq:Tcoll}
c T_S^{\mathrm{coll}}  = 
\bigintss _{0} ^{r_i} 
\dfrac{  \sqrt{r_i-2m} \, \sqrt{r^3} \, dr}{(r-2m) \sqrt{2m(r_i-r)}}
- \bigintss _{0}^{r_i} \dfrac{\sqrt{2mr} \, dr}{(r-2m)} \, .
\end{equation}
Note that necessarily $2m<r_i$. If $2m<r_i \le 3m$, the star casts the same shadow
as an eternal black hole, for any observer position outside the star. The reason is 
that then the past-oriented light rays from the observer position separate into
the same two classes as in the case of an eternal black hole: there is the class of 
light rays that go to infinity and the class of light rays that do not, with the 
borderline corresponding to light rays that asymptotically spiral towards the
light sphere at $r=3m$. So the formulas of the preding section apply to this case 
as well. Of course, here it is crucial that the star is assumed to be dark and
non-transparent.

We will, thus, assume from now on that the star collapses from an initial 
radius $r_i> 3m$. 
For calculating the shadow we have to consider lighlike geodesics
that graze the surface of the collapsing star. If such a lightlike
geodesic passes through a minimum radius value $r_m$, we may
determine $r_m$ by equating the first and the last expression in 
(\ref{eq:alpha2}) with $r=r_S$, $\tilde{\alpha} = \pi/2$ and $\varepsilon ^2 = 
1- 2m/r_i$. This results in

\begin{equation}\label{eq:rmcoll2}
\dfrac{r_m^3}{r_m-2m} = \dfrac{r_i r_S^2}{r_i-2m} \, . 
\end{equation}
Recall that a minimum value is possible only for $r_m>3m$,
i.e., in (\ref{eq:rmcoll2}) $r_S$ must satisfy the inequality $r_S>r_S^{(2)}$
where 

\begin{equation}\label{eq:rS2}
r_S ^{(2)} = 3 m \sqrt{3-\dfrac{6m}{r_i}} \, .
\end{equation}
As $r_i>3m$, (\ref{eq:rS2}) implies that 

\begin{equation}\label{eq:rS2in}
3m<r_S^{(2)}<r_i \, .
\end{equation}
If $r_i$ varies over its allowed values from $3m$ to infinity, $r_S^{(2)}$ 
monotonically increases from $3m$ to $\sqrt{27} \, m$. 

By (\ref{eq:surface}), the star passes through the critical radius value 
$r_S^{(2)}$ at
 
\begin{equation}\label{eq:TS2}
c T_S  ^{(2)}= 
\bigintss _{r_S^{(2)}} ^{r_i} 
\dfrac{  \sqrt{r_i-2m} \, \sqrt{r^3} \, dr}{(r-2m) \sqrt{2m(r_i-r)}}
- \bigintss _{r_S^{(2)}}^{r_i} \dfrac{\sqrt{2mr} \, dr}{(r-2m)} \, .
\end{equation}

We divide the collapse of the star into three phases, see again 
Figure~\ref{fig:collstat}: In the first phase from $T_S= - \infty$ 
to $T_S=0$ the star has a constant radius $r_i$. In the second phase
from $T_S=0$ to $T_S=T_S^{(2)}$ the star collapses to the 
critical radius value $r_S^{(2)}$. In the third phase from $T_S=T_S^{(2)}$
to $T_S=T_S^{\mathrm{coll}}$ the star completes the collapse.

In Figure~\ref{fig:collstat} we have indicated the worldline of an
observer who is static at radius coordinate $r_O$. We will now 
discuss the shadow of the collapsing star as seen by this observer.
As necessarily $2m< r_O$, we have to distinguish the following three
cases, in accordance with (\ref{eq:rS2in}): 
(a) $r_i<r_O$, (b) $r_S^{(2)} < r_O < r_i$ and (c) $2m < r_O
< r_S^{(2)}$.

In case (a), we distinguish three phases of the development of the 
shadow, corresponding to the three phases of the collapse.  In the first
phase the observer sees a static star of radius $r_i$. As the star is assumed 
to be dark, the observer sees a shadow whose angular radius is determined 
by light rays grazing the surface of the star, i.e., by light rays going through
a minimum of the radius coordinate at $r_m=r_i$. From (\ref{eq:alpha4})
we read that the angular radius $\alpha _{\mathrm{sh}}$ of this shadow is 
given by

\begin{equation}\label{eq:shcollstat1}
\mathrm{sin} \, \alpha _{\mathrm{sh}} =
\sqrt{\dfrac{r_i^3(r_O-2m)}{r_O^3(r_i-2m)}}
\, .
\end{equation}
This first phase ends when the observer sees the beginning of the collapse, i.e.,
at an observer time $T_O^{(1)}$ when a light signal that has gone 
through its minimum radius value $r_m=r_i$  at time $T_S=0$ reaches
the observer at $r_O$. From (\ref{eq:trtime2}) with the plus sign we find 
that 

\begin{equation}\label{eq:TO1}
c \, T_O ^{(1)}  = 
\bigintss _{r_i} ^{r_O} 
\dfrac{\sqrt{2mr} \, dr}{(r-2m)}
+
 \bigintss _{r_i}^{r_O} \dfrac{\sqrt{(r_i-2m)r^5} \, dr}{(r-2m) \sqrt{(r_i-2m)r^3-(r-2m)r_i^3}}
\, .
\end{equation}
During the second phase the observer sees a collapsing star. The boundary of the shadow 
is determined by light rays that graze the surface of the \emph{collapsing} star.
The minimum radius value $r_m$ of such light rays is given by (\ref{eq:rmcoll2}).
Inserting this value into (\ref{eq:alpha4}), with $r=r_O$, gives 
us the angular radius of the shadow in the second phase as a  
function of the parameter $r_S$,

\begin{equation}\label{eq:shcollstat2}
\mathrm{sin} \, \alpha _{\mathrm{sh}}=
\sqrt{\dfrac{r_i r_S^2(r_O-2m)}{r_O^3(r_i-2m)}} \, .
\end{equation}

\begin{figure}[h]
    \psfrag{x}{$c T _O$} % 
    \psfrag{y}{$\,$ \hspace{-0.3cm} $\alpha _{\mathrm{sh}}$} % 
    \psfrag{a}{\small $\,$ \hspace{-0.64cm} $\begin{matrix} \, \\[-0.12cm] 15 m\end{matrix}$} % 
    \psfrag{b}{\small $\,$ \hspace{-0.6cm} $\begin{matrix} \, \\[-0.12cm] 30 m\end{matrix}$} % 
    \psfrag{p}{\small $\,$ \hspace{-0.55cm} $\begin{matrix} \, \\[-0.2cm] cT_O^{(1)} \end{matrix}$} % 
    \psfrag{q}{\small $\,$ \hspace{-0.6cm} $\begin{matrix} \, \\[-0.2cm] cT_O^{(2)} \end{matrix}$} % 
    \psfrag{f}{\small $\,$ \hspace{-0.5cm} $\dfrac{\pi}{4}$} % 
    \psfrag{e}{\small $\,$ \hspace{-0.5cm} $\dfrac{\pi}{8}$} % 
\centerline{\epsfig{figure=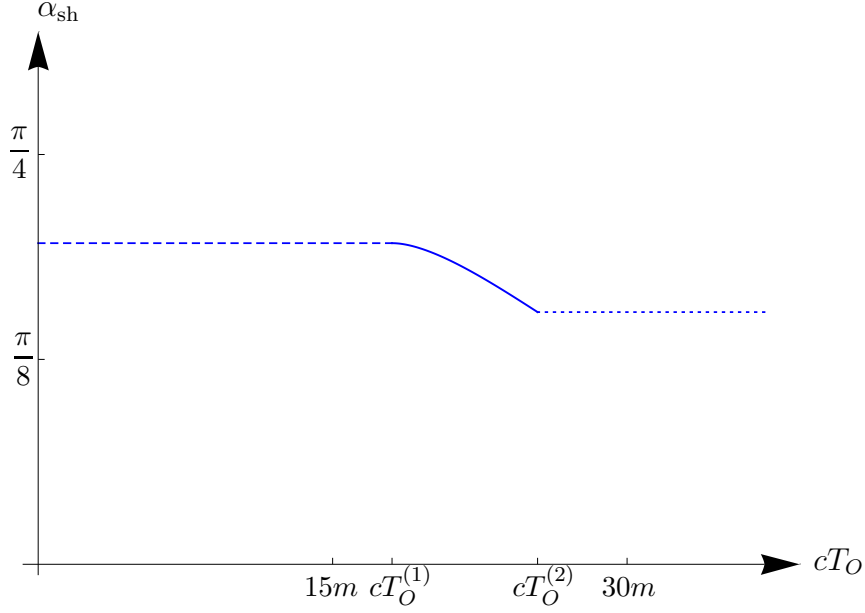, width=10.7cm}}
\caption{Angular radius $\alpha _{\mathrm{sh}}$ of the shadow of a collapsing 
dark star for a static observer. The observer is at $r_O=10 m$, the 
surface of the star is assumed to collapse from $r_i = 5m$.
As a function of the Painlev{\'e}-Gullstrand time coordinate 
$T _O$ of the observer, $\alpha _{\mathrm{sh}}$
is constant until $T_O^{(1)}$ which is shown as a dashed line, then it decreases 
according to the parametric description given by (\protect\ref{eq:Tcollstat})
and (\protect\ref{eq:shcollstat2}) until time $T_O^{(2)}$, and
from $T_O^{(2)}$ on it is given by Synge's formula 
(\protect{\ref{eq:alpha6}}) which is shown as a dotted line. \label{fig:alphacoll1}}
\end{figure}

The time $T_O$ at which the shadow with this angular radius is seen is found
by integrating  (\ref{eq:trtime2}),

\begin{equation}\label{eq:trcollstat}
c (T_O-T_S )  = 
\bigintss _{r_S} ^{r_O} 
\dfrac{\sqrt{2mr} \, dr}{(r-2m)}
+
 \bigintss _{r_S}^{r_O} \dfrac{\sqrt{r_i-2m} \, \sqrt{r^5} \, dr}{(r-2m) \sqrt{(r_i-2m)r^3-(r-2m)r_ir_S^2}}
\end{equation}
where again we have chosen the plus sign in (\ref{eq:trtime2}) because
$T_O>T_S$.  With (\ref{eq:surface}) this results in

\begin{gather}\label{eq:Tcollstat}
c T _O =    
\bigintss _{r_S} ^{r_i} 
\dfrac{  \sqrt{r_i-2m} \, \sqrt{r^3} \, dr}{(r-2m) \sqrt{2m(r_i-r)}}
+ \bigintss _{r_i} ^{r_O} 
\dfrac{\sqrt{2mr} \, dr}{(r-2m)}
\\
\nonumber
+
 \bigintss _{r_S}^{r_O} \dfrac{\sqrt{r_i-2m} \, \sqrt{r^5} \, dr}{(r-2m) \sqrt{(r_i-2m)r^3-(r-2m)r_ir_S^2}}
\, .
\end{gather}

If $r_i$ and $r_O$ are given, with $3m<r_i<r_O$, (\ref{eq:Tcollstat}) and 
(\ref{eq:shcollstat2}) give us the relation between $T _O$ and $\alpha _{\mathrm{sh}}$
in parametric form, $T _O = f_1 (r_S)$ and $\alpha _{\mathrm{sh}} = f_2 (r_S)$,
i.e., they give us the angular radius of the shadow in analytic form. This relation 
is valid in the second phase which lasts from $T_O^{(1)}$ up to a time
$T_O^{(2)}$. In this time interval, $r_S$ runs down from $r_S^{(1)}=r_i$ to the 
value $r_S^{(2)}$ given in (\ref{eq:rS2}). From (\ref{eq:Tcollstat}) we find that

\begin{gather}\label{eq:TO2}
c T _O ^{(2)}=    
\bigintss _{r_S^{(2)}} ^{r_i} 
\dfrac{  \sqrt{r_i-2m} \, \sqrt{r^3} \, dr}{(r-2m) \sqrt{2m(r_i-r)}}
+ \bigintss _{r_i} ^{r_O} 
\dfrac{\sqrt{2mr} \, dr}{(r-2m)}
\\
\nonumber
+
 \bigintss _{r_S^{(2)}}^{r_O} 
\dfrac{\sqrt{r^5} \, dr}{(r-2m) \sqrt{r^3-(r-2m)27m^2}}
\, .
\end{gather}

In the third phase, i.e., for times $T_O > T_O^{(2)}$, the angular radius of 
the shadow is given by Synge's formula (\ref{eq:alpha6}). Past-oriented light 
rays grazing the surface of the star cannot escape to infinity anymore,
i.e., they do not give the boundary of the shadow; the latter is determined by light
rays that spiral asymptotically to $r=3m$.

\begin{figure}[h]
    \psfrag{x}{$r_O$} % 
    \psfrag{y}{$\,$ \hspace{-1.3cm} $cT_O^{(2)}-cT_O^{(1)}$} % 
    \psfrag{a}{\small $\,$ \hspace{-0.5cm} $25m$} % 
    \psfrag{b}{\small $\,$ \hspace{-0.5cm} $50m$} % 
    \psfrag{c}{\small $\,$ \hspace{-0.5cm} $75m$} % 
    \psfrag{d}{\small $\,$ \hspace{-0.5cm} $100m$} % 
    \psfrag{e}{\small $\,$ \hspace{-0.9cm} $25m$} % 
    \psfrag{f}{\small $\,$ \hspace{-0.9cm} $50m$} % 
    \psfrag{g}{\small $\,$ \hspace{-0.9cm} $75m$} % 
\centerline{\epsfig{figure=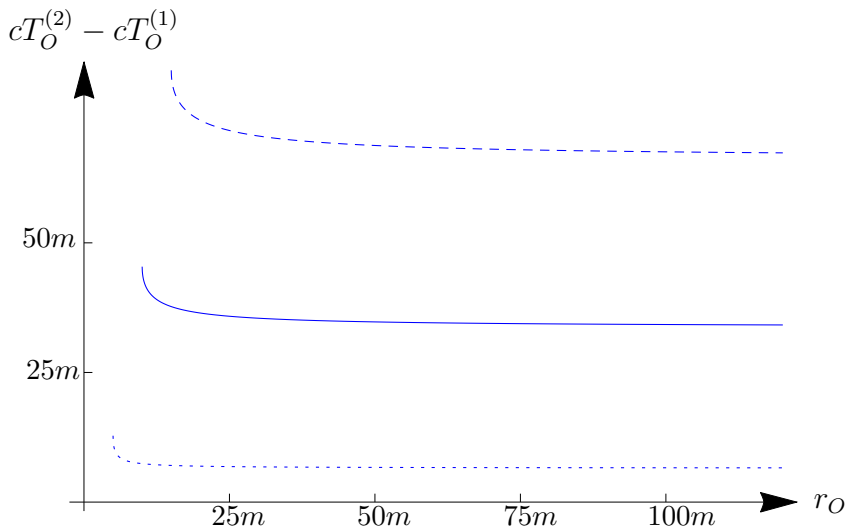, width=10.1cm}}
\caption{Time $T_O^{(2)}-T_O^{(1)}$ over which the static observer sees the 
star collapse, plotted against the observer position $r_O$. The star is
collapsing from an initial radius $r_i=5m$ (dotted), $r_i=10m$ (solid)
or $r_i=15m$ (dashed), respectively. \label{fig:colltime}}
\end{figure}

We summarise our analysis in the following way. In the first phase, which lasts 
from $T_O = - \infty$ to $T_O = T_O^{(1)}$ given by (\ref{eq:TO1}), the observer
sees a shadow of constant angular radius given by (\ref{eq:shcollstat1}). In the
second phase, which lasts from $T_O= T_O^{(1)}$ until $T_O=T_O^{(2)}$ given by
(\ref{eq:TO2}), the observer sees a shrinking shadow whose angular radius
as a function of observer time $T_O$ is given in parametric form by (\ref{eq:Tcollstat})
and (\ref{eq:shcollstat2}). The parameter $r_S$ runs down from $r_S^{(1)}=r_i$ to
$r_S^{(2)}= 3m \sqrt{3-6m/r_i}$. The third phase lasts from $T_O = T_O^{(2)}$ to
$T_O=\infty$. In this period the observer sees a shadow of constant angular
radius given by Synge's formula (\ref{eq:alpha6}). The angular radius  of the
shadow is plotted against $T_O$, over all three periods, for $r_i=5m$
and $r_O=10m$ in Figure \ref{fig:alphacoll1}. 

In Fig. \ref{fig:colltime} we plot the time $T_O^{(2)}-T_O^{(1)}$ over which 
the observer sees the star collapse against the observer position $r_O$. We
see that this time is largely independent of $r_O$, unless the observer is
very close to the star. For a star collapsing from an initial radius of 5
Schwarzschild radii, $r_i=10m$, we see that $T_O^{(2)}-T_O^{(1)} \approx
34 \, m/c$ for a sufficiently distant observer. For a stellar black hole, a typical
value would be $m \approx 15 \, \mathrm{km}$, resulting in $T_O^{(2)}-T_O^{(1)}
\approx 0.001 \, \mathrm{sec}$, so such a collapse would happen quite quickly.
Even for a supermassive black hole of $m \approx 10^6 \, \mathrm{km}$, the
observer would see the collapse happen in less than 2 minutes. For the case 
of a collapsing cluster of galaxies the formation of the shadow would take longer, 
but in this case it is more reasonable to model the collapsing object as transparent.
Note that on the worldline of a distant static observer Painlev{\'e}-Gullstrand 
time $T_O$ is practically the same as proper time $\tau _O$ because, 
by (\ref{eq:taustatic}), 
 
\begin{equation}\label{eq:tauTO}
 \tau _O  = \sqrt{1- \dfrac{2m}{r_O}} \,  T_O + \mathrm{constant}.
\end{equation}

\begin{figure}[h]
    \psfrag{x}{$c T _O$} % 
    \psfrag{y}{$\,$ \hspace{-0.3cm} $\alpha _{\mathrm{sh}}$} % 
    \psfrag{a}{\small $\,$ \hspace{-0.62cm} $\begin{matrix} \, \\[-0.12cm] 4 m\end{matrix}$} % 
    \psfrag{b}{\small $\,$ \hspace{-0.58cm} $\begin{matrix} \, \\[-0.12cm] 8 m\end{matrix}$} % 
    \psfrag{c}{\small $\,$ \hspace{-0.62cm} $\begin{matrix} \, \\[-0.12cm] 12 m\end{matrix}$} % 
    \psfrag{d}{\small $\,$ \hspace{-0.58cm} $\begin{matrix} \, \\[-0.12cm] 16 m\end{matrix}$} % 
    \psfrag{p}{\small $\,$ \hspace{-0.56cm} $\begin{matrix} \, \\[-0.2cm] cT_O^{(0)} \end{matrix}$} % 
    \psfrag{q}{\small $\,$ \hspace{-0.64cm} $\begin{matrix} \, \\[-0.2cm] cT_O^{(2)} \end{matrix}$} % 
    \psfrag{e}{\small $\,$ \hspace{-0.5cm} $\dfrac{\pi}{4}$} % 
    \psfrag{f}{\small $\,$ \hspace{-0.5cm} $\dfrac{\pi}{2}$} % 
    \psfrag{r}{\small $\,$ \hspace{-0.85cm} $\alpha _{\mathrm{sh}}^{(0)}$} % 
\centerline{\epsfig{figure=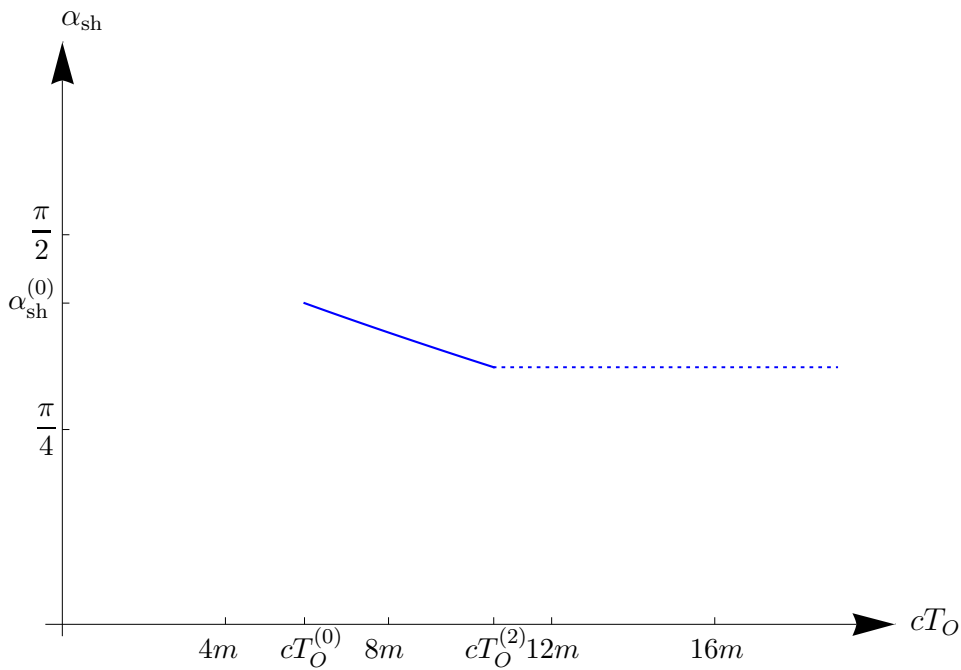, width=11.7cm}}
\caption{Angular radius $\alpha _{\mathrm{sh}}$ of the shadow of a collapsing 
dark star for a static observer. The observer is at $r_O=4.5 m$, the 
surface of the star is collapsing from $r_i = 5 m$.
The observation begins at Painlev{\'e}-Gullstrand time $T_O^{(0)}$ 
when the surface of the star passes through the radius value $r_O$. 
At this moment the angular radius of the shadow takes a value $\alpha _{\mathrm{sh}}^{(0)}$ 
which is smaller than $\pi /2$, because of aberration.
As a function of the Painlev{\'e}-Gullstrand  time coordinate $T _O$, the 
angular radius $\alpha _{\mathrm{sh}}$ then decreases until it becomes a 
constant at time $T_O^{(2)}$. This constant value, which is again shown as a dotted
line, is given by Synge's formula (\protect{\ref{eq:alpha6}}). \label{fig:alphacoll2}}
\end{figure}

A very similar analysis applies to case (b). The only difference is that then in 
the beginning the observer is inside the star. The observation can begin only at 
the time when the surface of the star passes through the radius value $r_O$
which, by assumption, is bigger than $r_S^{(2)}$. From that time on, the angular 
radius of the shadow is given by the same equations as before for the second and 
the third phase. A plot of the angular radius of the shadow against $T_O$ is
shown in Figure~\ref{fig:alphacoll2} for $r_O=4.5 m$ and $r_i=5m$.

In case (c) the observer is initially inside the star, as in case (b). The
difference is in the fact that now the radius of the star is smaller than $r_S^{(2)}$ 
at the moment when the observation begins. Therefore, the shadow is never
determined by light rays that graze the surface of the star; it is always 
determined by light rays that spiral towards $r=3m$, i.e., the angular radius 
of the shadow is constant from the beginning of the observation and given by 
Synge's formula.    
  
%%%%%%%%%%%%%%%%%%%%%%%%%%%%%%%%%%%%%%%%%%%%%%%%%%%%%%%%%%%%%%%%%%%%
\section{The shadow of a collapsing star for an infalling observer}\label{sec:shcoll2}

\begin{figure}[h]
    \psfrag{x}{$r $} % 
    \psfrag{y}{$\,$ \hspace{-0.3cm} $T$} % 
    \psfrag{z}{\small $\,$ \hspace{-0.45cm} $\begin{matrix} \, \\[-0.1cm] r_O^* \end{matrix}$} % 
    \psfrag{a}{\small $\,$ \hspace{-0.35cm} $\begin{matrix} \, \\[-0.03cm] r_i \end{matrix}$} % 
    \psfrag{b}{\small $\,$ \hspace{-0.6cm} $\begin{matrix} \, \\[-0.1cm] r_S^{(2)} \end{matrix}$} % 
    \psfrag{p}{\small $\,$ \hspace{-0.8cm} $T_S^{(2)}$} % 
    \psfrag{q}{\small $\,$ \hspace{-0.9cm} $T_S^{\mathrm{coll}}$} % 
\centerline{\epsfig{figure=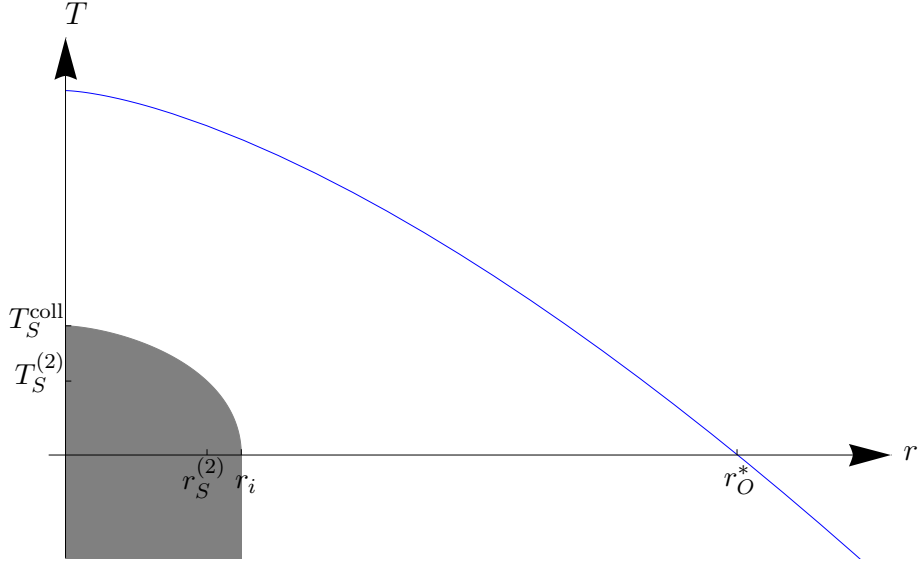, width=11.5cm}}
\caption{Spacetime diagram of a collapsing star and an infalling
observer. The star begins to collapse at Painlev{\'e}-Gullstrand time
$T=0$ with radius $r_i$. The freely falling observer passes at this time through
the radius value $r_O^*$. \label{fig:collcoll}}
\end{figure}

We consider the same collapsing dark star as in the preceding section,
but now we want to calculate the shadow as it is seen by an infalling
observer. The relation between the coordinates $r_O$ and $T_O$ of the 
infalling observer can be found by integrating (\ref{eq:rTrad}),
\begin{equation}\label{eq:TOrO}
c T_O  = \bigintss_{r_O} ^{r_O^*} 
\dfrac{
\Big( \varepsilon  \sqrt{r^3}- \sqrt{2mr} \sqrt{ \varepsilon  ^2 r -r+2m}\Big) dr
}{
( r -2m ) \sqrt{\varepsilon  ^2r -r+2m} 
} \, .
\end{equation}
Here $r_O^*$ is an integration constant that gives the position of the observer at $T=0$ which
is the time when the star begins to collapse.
We assume that $r_i>3m$, hence $3m < r_S^{(2)}<r_i$, and that $r_O^*$ has been chosen 
big enough such that the observer is outside the star for all times, see Figure~\ref{fig:collcoll}.
For the time being we leave the constant of motion $\varepsilon$ unspecified.

\vspace{0.5cm}

We will determine the angular radius of the shadow as a function of the observer position 
$r_O$. As before, we distinguish three phases. In the first phase the observer
sees a star of constant radius $r_i$. The angular radius of the shadow can be read
from (\ref{eq:alpha5}) with the lower sign where we have to insert $r=r_O$ and $r_m=r_i$,

\begin{equation}\label{eq:shcollcoll1}
\mathrm{sin} \, \tilde{\alpha}{}_{\mathrm{sh}} =
\dfrac{
(r_O -2m) \, \sqrt{r_i^3} 
}{
\varepsilon \, r_O^2 \, \sqrt{r_i-2m} - \, 
\sqrt{\varepsilon ^2 r_O -r_O+2m} \, 
\sqrt{(r_i-2m) r_O^3-(r_O-2m)r_i^3}
} \, .
\end{equation}
If $r_i$ is given, this gives us explicitly $\tilde{\alpha}{}_{\mathrm{sh}}$ as 
a function of $r_O$ for the first phase.

In the second phase  we may again use (\ref{eq:Tcollstat}). In combination
with (\ref{eq:TOrO}) this implies

\begin{gather}\label{eq:arrivaltime2}
\bigintss _{r_O} ^{r_O^*} 
\dfrac{\varepsilon \, \sqrt{r^3} \, dr}{(r-2m) \, \sqrt{\varepsilon ^2 r - r +2m}}
-
\bigintss _{r_i} ^{r_O^*} 
\dfrac{\sqrt{2mr} \, dr}{(r-2m)}
\\
\nonumber
=
 \bigintss _{r_S}^{r_i} 
\dfrac{\sqrt{r_i-2m} \, \sqrt{r^5} \, dr}{(r-2m) \sqrt{2m(r_i-r)}}
+
\bigintss_{r_S} ^{r_O} 
\dfrac{
 \sqrt{r_i-2m} \, \sqrt{r^5} \,  dr
}{
(r-2m) \sqrt{(ri-2m)r^3-(r-2m)r_ir_S^2} 
}
\, .
\end{gather}
The angular radius of the shadow is again given by (\ref{eq:alpha5}) 
with the lower sign where now we have to insert $r=r_O$ and $r_m$
from (\ref{eq:rmcoll2}),

\begin{equation}\label{eq:shcollcoll2}
\mathrm{sin} \, \tilde{\alpha}{}_{\mathrm{sh}} =
\dfrac{
(r_O -2m) \, r_S \, \sqrt{r_i} 
}{
\varepsilon \, r_O^2 \, \sqrt{r_i-2m} - \, 
\sqrt{\varepsilon ^2 r_O-r_O+2m} \, 
\sqrt{(r_i-2m) r_O^3-(r_O-2m)r_i r_S^2}
} \, .
\end{equation}

\begin{figure}[h]
    \psfrag{x}{$r _O$} % 
    \psfrag{y}{$\,$ \hspace{-0.3cm} $\tilde{\alpha}{} _{\mathrm{sh}}$} % 
    \psfrag{a}{\small $\,$ \hspace{-0.62cm} $\begin{matrix} \, \\[0.07cm] 3 m\end{matrix}$} % 
    \psfrag{b}{\small $\,$ \hspace{-0.56cm} $\begin{matrix} \, \\[0.053cm] r_O^{(2)} \end{matrix}$} % 
    \psfrag{c}{\small $\,$ \hspace{-0.62cm} $\begin{matrix} \, \\[0.053cm] r _O^{(1)} \end{matrix}$} % 
    \psfrag{d}{\small $\,$ \hspace{-0.58cm} $\begin{matrix} \, \\[0.07cm] 10 m\end{matrix}$} % 
    \psfrag{e}{\small $\,$ \hspace{-0.58cm} $\dfrac{\pi}{2}$} % 
\centerline{\epsfig{figure=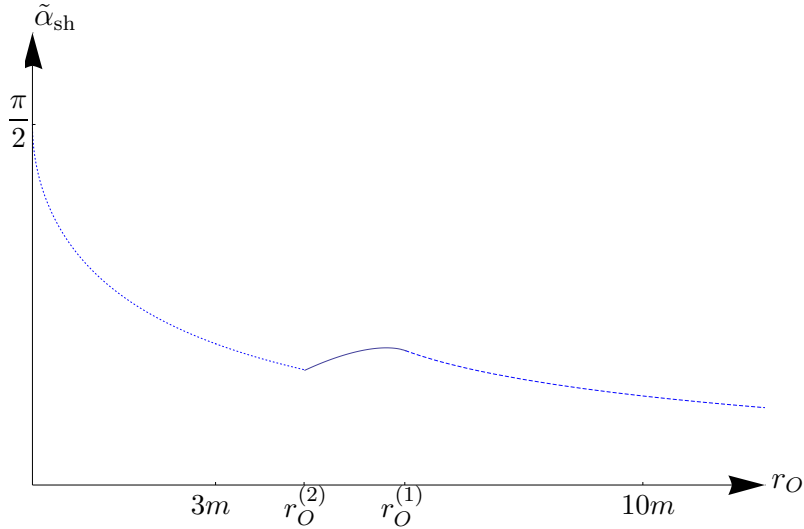, width=10cm}}
\caption{Angular radius $\tilde{\alpha}{} _{\mathrm{sh}}$ of the shadow of a collapsing 
dark star for an infalling observer. The observer is a Painlev{\'e}-Gullstrand observer ($\varepsilon = 1$)
passing at $T_O=0$ through the radius value $r_O^*=10 \, m$. At this time the 
surface of the star starts collapsing from $r_i = 5 m$.
The angular radius of the shadow is plotted against the radius coordinate 
of the observer. We distinguish three phases: In the first phase (dashed), the observer 
sees a star of constant radius $r_i$ and the angular radius of the shadow is given 
by (\protect{\ref{eq:shcollcoll1}}). In the second phase (solid), the observer sees a 
collapsing star and the angular radius of the shadow is implicitly given by 
(\protect{\ref{eq:arrivaltime2}}) with $r_S$ inserted from (\protect{\ref{eq:rScollcoll}}).
In the third phase (dotted), the boundary of the shadow is no longer given by 
light rays grazing the surface of the star but rather by light rays spiralling
towards the photon sphere at $r=3m$, so $\tilde{\alpha}{}_{\mathrm{sh}}$ is
given by (\protect{\ref{eq:alpha7}}), cf. Fig.~\protect{\ref{fig:talphabh}}. \label{fig:alphacoll3}}
\end{figure}

This equation can be solved for $r_S$,

\begin{equation}\label{eq:rScollcoll}
r_S = 
\dfrac{
\sqrt{r_i-2m} \, \sqrt{r_O^3} \, \mathrm{sin} \, \tilde{\alpha}{}_{\mathrm{sh}}
}{
\sqrt{r_i} \Big( 
\varepsilon \, \sqrt{r_O} - \sqrt{\varepsilon ^2 r_O - r_O +2m} \,
\mathrm{cos} \, \tilde{\alpha}{}_{\mathrm{sh}} \Big)
} \, .
\end{equation}
Inserting (\ref{eq:rScollcoll}) into (\ref{eq:arrivaltime2}) gives us the desired
relation between $r_O$ and $\tilde{\alpha}{}_{\mathrm{sh}}$ in implicit but
fully analytical form, provided that $r_i$ and $r_O^*$ are prescribed. The
second phase begins when the observer passes
through a radius value $r_O=r_O^{(1)}$ such
that (\ref{eq:arrivaltime2}) holds with $r_S=r_i$. It ends at a radius
value  $r_O=r_O^{(2)}$ such that (\ref{eq:arrivaltime2}) holds with $r_S=3m$.    

Finally, in the third phase the boundary of the shadow is determined
by light rays that spiral asymptotically to $r=3m$, i.e., $\tilde{\alpha}{}_{\mathrm{sh}}$
is given by (\ref{eq:alpha7}).

%%%%%%%%%%%%%%%%%%%%%%%%%%%%%%%%%%%%%%%%%%%%%%%%
\section{Conclusions}\label{sec:conclusions}
In this paper we have demonstrated that, for a spherically symmetric 
dark and non-transparent star that collapses in free fall like a ball of dust, 
the development of the shadow can be calculated analytically, both for a 
static and for an infalling observer. In particular we have shown that for a 
static observer the black-hole shadow according to Synge's formula
forms in a finite time which, for a stellar black hole, is in the order of 
fractions of a second. This result could not have been easily anticipated 
before doing the calculation: Intuitively, one might have expected that 
the black-hole shadow forms asymptotically. The situation is similar
for an infalling observer (provided that the observer is sufficiently
far behind not to catch up with the star):
Also in this case the surface of the star determines the shadow 
only over a finite time; during the last stage of the infall, the 
observer sees the same shadow as when infalling into an eternal 
black hole.     

Admittedly, getting analytical results was possible only because we
used a somewhat oversimplified model for a collapsing star. More 
realistically, instead of a spherically symmetric ball of dust one should 
consider a rotating star with pressure which would probably make the
calculations so complicated that only a numerical treatment would
be possible. However, we believe that the simple model considered
here gives a good idea of all the relevant qualitative features
of how the black-hole shadow comes about in the course of time.

In this paper we have concentrated on the
formation of the shadow during gravitational collapse. However, we
mention that some of our results may also be useful for investigating
the temporal change of the shadow of an already existing black
hole. If a black hole is surrounded by matter its mass will grow by
accretion, so its shadow will become bigger in the course of time.
We have not investigated this problem in detail, but we believe that
the Painlev{\'e}-Gullstrand approach pursued in this paper may be
appropriate also for calculating the growth of the shadow of an accreting
black hole.

%%%%%%%%%%%%%%%%%%%%%%%%%%%%%%%%%%%%%%%%%%%%%%%%%%%%%%%%%%%%%%%%%%%%%%%%%
\section*{Acknowledgements}
We would like to thank Nico Giulini for helpful discussions.
Moreover, we gratefully acknowledge support from the DFG 
within the Research Training Group 1620 ``Models of Gravity''.

\bibliographystyle{spphys}       % APS-like style for physics
%\bibliography{ref} 

\begin{thebibliography}{10}
\providecommand{\url}[1]{{#1}}
\providecommand{\urlprefix}{URL }
\expandafter\ifx\csname urlstyle\endcsname\relax
  \providecommand{\doi}[1]{DOI \discretionary{}{}{}#1}\else
  \providecommand{\doi}{DOI \discretionary{}{}{}\begingroup
  \urlstyle{rm}\Url}\fi

\bibitem{Synge1966}
J. L. Synge, The escape of photons from gravitationally intense stars, Mon. Not.
  Roy. Astron. Soc. \textbf{131}, 463 (1966)

\bibitem{Bardeen1973}
J.~Bardeen, in \emph{Black Holes}, ed. by C.~DeWitt, B.~DeWitt (Gordon and
  Breach, New York, U.S.A., 1973), p. 215

\bibitem{GrenzebachPerlickLaemmerzahl2014}
A.~Grenzebach, V.~Perlick, C.~L{\"a}mmerzahl, Photon regions and shadows of
  {K}err-{N}ewman-{NUT} black holes with a cosmological constant, Phys. Rev.
  \textbf{D 89}, 124004 (2014)

\bibitem{GrenzebachPerlickLaemmerzahl2015}
A.~Grenzebach, V.~Perlick, C.~L{\"a}mmerzahl, Photon regions and shadows of
  accelerated black holes, Int. J. Modern Phys. \textbf{D 24}, 1542024 (2015)

\bibitem{Tsupko2017}
O. Yu. Tsupko, Analytical calculation of black hole spin using deformation of the shadow, Phys. Rev. \textbf{D 95}, 104058 (2017)

\bibitem{FalckeMeliaAgol2000}
H.~Falcke, F.~Melia, E.~Agol, Viewing the shadow of the black hole at the
  galactic center, Astrophys. J. \textbf{528}, L13 (2000)

\bibitem{JamesTunzelmannFranklinThorne2015}
O.~James, E.~Tunzelmann, P.~Franklin, K.~Thorne, Gravitational lensing by
  spinning black holes in astrophysics, and in the movie \emph{Interstellar},
  Class. Quant. Grav. \textbf{32}, 065001 (29015)

\bibitem{AmesThorne1968}
W.~Ames, K.~Thorne, The optical appearance of a star that is collapsing through
  its gravitational radius, Astrophys. J. \textbf{151}, 659 (1968)

\bibitem{Jaffe1969}
J.~Jaffe, Collapsing objects and the backward emission of light, 
Ann. Phys. (NY) \textbf{55}, 374 (1969)

\bibitem{LakeRoeder1979}
K.~Lake, R.~C.~Roeder, Note on the optical appearance of a star 
collapsing through its gravitational radius, 
Astrophys. J. \textbf{232}, 277 (1979)
	
\bibitem{FrolovKimLee2007}
V.~P.~Frolov, K.~Kim, H.~K.~Lee, Spectral broadening of radiation from 
relativistic collapsing objects, 
Phys. Rev. \textbf{D 75}, 087501 (2007)

\bibitem{KongMalafarinaBambi2014}
L.~Kong, D.~Malafarina, C.~Bambi,
Can we observationally test the weak cosmic censorship conjecture?
Eur. Phys. J. {\textbf C 74} 2983 (2014)

\bibitem{KongMalafarinaBambi2015}
L.~Kong, D.~Malafarina, C.~Bambi,
Gravitational blueshift from a collapsing object,
Phys. Lett. {\textbf B 741} 82 (2015)

\bibitem{OrtizSarbachZannias2015a}
N.~Ortiz, O.~Sarbach, T.~Zannias, Shadow of a naked singularity \textbf{92},
  044035 (2015)

\bibitem{OrtizSarbachZannias2015b}
N.~Ortiz, O.~Sarbach, T.~Zannias, Observational distinction between black holes
  and naked singularities: the role of the redshift function, Class. Quant.
  Grav. \textbf{32}, 247001 (2015)

\bibitem{OppenheimerSnyder1939}
J. R. Oppenheimer, H.~Snyder, On continued gravitational contraction, Phys. Rev.
  \textbf{56}, 455 (1939)

\bibitem{Painleve1921}
P.~Painlev{\'e}, La m{\'e}canique classique et la th{\'e}orie de la
  relativit{\'e}, C. R. Acad. Sci. \textbf{173}, 677 (1921)

\bibitem{Gullstrand1922}
A.~Gullstrand, Allgemeine {L}{\"o}sung des statischen {E}ink{\"o}rperproblems
  in der {E}insteinschen {G}ravitationstheorie, Ark. Mat. Astr. Fys.
  \textbf{16}, 1 (1922)

\bibitem{Lemaitre1933}
G.~Lema{\^\i}tre, L'{U}nivers en expansion, Ann. Soc. Sci. Bruxelles \textbf{A
  53}, 51 (1933)

\bibitem{BakalaEtAl2007}
P.~Bakala, P.~{\v C}erm{\'a}k, S.~Hled{\'\i}k, Z.~Stuchl{\'\i}k,
  K.~Truparov{\'a}, Extreme gravitational lensing in vicinity of
  {S}chwarzschild-de{S}itter black holes, Centr. Eur. J. Phys. \textbf{5},
  599 (2007)

\end{thebibliography}

\end{document}